\documentclass{article}
\usepackage{amssymb}
\usepackage{amsmath}
\usepackage{amsfonts}
\usepackage{sw20ssur}
\usepackage{cite}
\usepackage{afterpage}

\input{tcilatex}
\begin{document}

\title{Non-factorizable joint probabilities and evolutionarily stable
strategies\\
in the quantum prisoner's dilemma game}
\author{Azhar Iqbal$^{\text{a,b}}$ and Derek Abbott$^{\text{a}}$ \\
$^{\text{a}}${\small School of Electrical \& Electronic Engineering, The
University of Adelaide, SA 5005, Australia.}\\
$^{\text{b}}${\small Centre for Advanced Mathematics and Physics,} {\small %
National University of Sciences \& Technology,}\\
{\small Campus of College of Electrical \& Mechanical Engineering, Peshawar
Road, Rawalpindi, Pakistan.}}
\maketitle

\begin{abstract}
The well known refinement of the Nash Equilibrium (NE) called an
Evolutionarily Stable Strategy (ESS) is investigated in the quantum
Prisoner's Dilemma (PD) game that is played using an Einstein-Podolsky-Rosen
type setting. Earlier results report that in this scheme the classical NE
remains intact as the unique solution of the quantum PD game. In contrast,
we show here that interestingly in this scheme a non-classical solution for
the ESS emerges for the quantum PD.
\end{abstract}


\section{Introduction}

In the area of quantum games~\cite%
{Mermin,Mermin1,MeyerDavid,MeyerDavid1,EWL,Vaidman,Brandt,BenjaminHayden,
Johnson,Johnson1,MarinattoWeber,IqbalToor1,Du,DuLi,IqbalToor2,Piotrowski,
IqbalToor3,Piotrowski1,FlitneyAbbottA,IqbalToor4,FlitneyAbbottC,NawazToor,
Shimamura,IqbalToor5,IchikawaTsutsui,CheonIqbal,Ozdemir,IqbalCheon,Ramzan,
FlitneyHollenberg1,IqbalCheonM,Bleiler1,IqbalCheonChapter,Ichikawa,
IqbalCheonAbbott,IqbalAbbott,FlitneyRecent,Bleiler2}, a result from a recent
paper \cite{IqbalCheon} shows that, in the quantization scheme based on
performing generalized Einstein-Podolsky-Rosen-Bohm (EPR-Bohm) experiments 
\cite{EPR,Bohm,Bell,Aspect,Peres,Cereceda}, the two-player quantum game of
Prisoner's Dilemma (PD) does not offer a new Nash Equilibrium (NE\footnote{%
In the rest of this paper we use NE to mean Nash Equilibrium or Nash
Equilibria. The correct meaning is judged from the context.}) \cite%
{Binmore,Rasmusen}, which is different from the classical NE of the game in
which both players play the strategy of defection. This quantization scheme
constructs the quantum PD in two steps:

1) The players' payoff relations are re-expressed in terms of joint
probabilities corresponding to generalized EPR-Bohm experiments involving a
bipartite system shared between two players. In a run each player receives
one part of the system while having two observables both of which are
dichotomic. A player's strategy is defined to be entirely classical that
consists of a linear combination (with real and normalized coefficients) of
choosing between his/her two observables. The scheme embeds the classical
game within the quantum game by placing constraints on joint probabilities.
These constraints guarantee that for factorizable joint probabilities the
classical game emerges along with its particular outcome.

2) As a set of joint probabilities that violates Bell's inequality must
always be non-factorizable, the corresponding quantum game is constructed by
retaining the constraints on joint probabilities, obtained in the last step,
while they can now be non-factorizable.

By constructing quantum games from non-factorizable joint probabilities,
which a quantum mechanical apparatus can provide, this quantization scheme
avoids state vectors and brings out the essence of quantum games without
referring to quantum mechanics---an important consideration in developing
the present approach to quantum games. Game theory finds applications in a
range of disciplines~\cite{AbbottShaliziDavies} and we believe that more
accessible approaches to quantum games remain in need of development.

It turns out that in this quantization scheme the constraints on joint
probabilities obtained for the game of PD, which embed the classical game
within the quantum game, come out to be so strong that the subsequent
permitting joint probabilities to become non-factorizable cannot change the
outcome of the game. The quantum PD game that is played in this framework,
therefore, generates an outcome identical to the one obtained in the
classical game in which both players defect. This finding motivates us in
the present paper to investigate if non-factorizable joint probabilities can
bring out some non-classical outcome for a refinement of the NE in the PD
game, while not affecting the NE itself.

In this paper we show that surprisingly this indeed is the case. That is,
with an EPR-Bohm type setting for playing a quantum game a set of
non-factorizable joint probabilities is able to produce a non-classical
outcome in the quantum PD game to a well known refinement on the set of
symmetric Nash equilibria---called an Evolutionarily Stable Strategy (ESS)~%
\cite{MaynardSmith,Weibull,Samuelson}. This contrasts interestingly with the
reported result~\cite{IqbalCheon} that for the same game, non-factorizable
joint probabilities are unable to produce a non-classical outcome for a NE
and the classical NE remains intact as the unique solution of the quantum PD
game.

Using the quantization schemes of Eisert et al.~\cite{EWL}\ and Marinatto
and Weber~\cite{MarinattoWeber}, the game-theoretic concept of an ESS was
originally investigated in the area of quantum games by Iqbal and Toor in a
series of papers~\cite{IqbalToor1, IqbalToor2, IqbalToor3, IqbalToor4,
IqbalToor5} and was reviewed by Iqbal and Cheon in a book chapter~\cite%
{IqbalCheonChapter}. The present paper addresses the issues raised in these
publications using the new approach towards constructing quantum games
recently proposed by Iqbal and Cheon~\cite{IqbalCheon},\ which exploits
non-factorizable property of quantum mechanical joint probabilities in the
construction of quantum games.

In a recent paper~\cite{IqbalAbbott} we have investigated a quantum version
of the Matching Pennies game played in this quantization scheme to find that
non-classical NE emerge in this game for sets of (quantum mechanical) joint
probabilities that maximally violate CHSH form of Bell's inequality~\cite%
{Peres}. The present paper considers the PD game in this quantization scheme
and explores the fate of a well known refinement of the NE concept in
relation to joint probabilities becoming non-factorizable.

\section{Evolutionarily Stable Strategy}

An ESS is the central solution concept of evolutionary game theory~\cite%
{Weibull,Samuelson} (EGT). In EGT genes are considered players in survival
games and players' strategies are the behavioral characteristics imparted by
genes to their host organism, while the payoff to a gene is the number of
offspring carrying that gene~\cite{Samuelson}. The players' strategies
(which the players genes play until the biological agents carrying those
genes die) and their payoffs become related as host organisms having
favourable behavioral characteristics are better able to reproduce than
others.

Referring to a pool of genes, the notion of an ESS considers a large
population of players (genes) in which players are matched in random
pair-wise contests. We call the two players in an interaction to be player $%
1 $ and player $2$. Each player can play the strategy $S$ or the strategy $%
S^{\prime }$ in a pair-wise interaction and the payoff matrix for the game
is given as

\begin{equation}
\begin{array}{c}
\text{Player }1%
\end{array}%
\begin{array}{c}
S \\ 
S^{\prime }%
\end{array}%
\overset{%
\begin{array}{c}
\text{Player }2%
\end{array}%
}{\overset{%
\begin{array}{ccc}
S &  & S^{\prime }%
\end{array}%
}{\left( 
\begin{array}{cc}
(a_{1},b_{1}) & (a_{2},b_{2}) \\ 
(a_{3},b_{3}) & (a_{4},b_{4})%
\end{array}%
\right) }},  \label{matrix}
\end{equation}%
where the two entries in the bracket are player $1$'s and player $2$'s
strategies, respectively. For example, player $1$'s payoff is $\Pi
_{1}(S,S)=a_{1}$ when both players play the strategy $S$. It is found useful
to define

\begin{equation}
\mathcal{A}{\small =}\left( 
\begin{array}{cc}
a_{1} & a_{2} \\ 
a_{3} & a_{4}%
\end{array}%
\right) {\small ,}\text{ }\mathcal{B}{\small =}\left( 
\begin{array}{cc}
b_{1} & b_{2} \\ 
b_{3} & b_{4}%
\end{array}%
\right)
\end{equation}%
to be player $1$'s and player $2$'s payoff matrices, respectively. We write
players' payoffs as $\Pi _{1,2}(x,y)$ where subscripts $1$ or $2$ refer to
the players and $x$ and $y$ in bracket are player $1$'s and player $2$'s
strategies, respectively.

An ESS deals with symmetric games in which

\begin{equation}
\Pi _{1}(x,y)=\Pi _{2}(y,x)\text{ and }\Pi _{1}(y,x)=\Pi _{2}(x,y)
\label{symmetric game}
\end{equation}%
saying that, for example, player $1$'s payoff when s/he plays $x$ and player 
$2$ plays $y$, is same as the player $2$'s payoff when s/he plays $y$ and
player $1$ plays $x$, where $x$ and $y$ can be either $S$ or $S^{\prime }$.
In words, in a symmetric game a player's payoff is determined by the
strategy, and not by the identity, of a player.

For a symmetric game using subscripts in payoff relations becomes redundant
as $\Pi (x,y)$ denotes payoff to an $x$-player against a $y$-player. This
allows not to refer to players at all and to describe $\Pi (x,y)$ as the
payoff to $x$-strategy against the $y$-strategy. The game given by the
matrix (\ref{matrix}) is symmetric when $\mathcal{A=B}^{T}$. The game of PD
is a symmetric game, which is defined by the constraint $%
a_{3}>a_{1}>a_{4}>a_{2}$.

Assume that, in random pair-wise contests, the strategy $x$ is played by a
number of players whose relative proportion in the population is $\epsilon $
whereas the rest of the population plays the strategy $x^{\star }$. EGT
defines the fitnesses \cite{Weibull,Samuelson} of the strategies $x$ and $%
x^{\star }$ as

\begin{equation}
F(x)=\epsilon \Pi (x,x)+(1-\epsilon )\Pi (x,y),\text{ }F(x^{\star
})=\epsilon \Pi (x^{\star },x)+(1-\epsilon )\Pi (x^{\star },x^{\star }),
\end{equation}%
in terms of which the strategy $x^{\star }$ is called an ESS when $%
F(x^{\star })>F(x)$ i.e.

\begin{equation}
\epsilon \Pi (x^{\star },x)+(1-\epsilon )\Pi (x^{\star },x^{\star
})>\epsilon \Pi (x,x)+(1-\epsilon )\Pi (x,x^{\star }).
\end{equation}%
Since $\epsilon \ll 1$, the terms containing $\epsilon $ can be ignored
effectively. So $F(x^{\star })>F(x)$ implies $\Pi (x^{\star },x^{\star
})>\Pi (x,x^{\star })$. If, however, $\Pi (x^{\star },x^{\star })=\Pi
(x,x^{\star })$, we need to consider the terms containing $\epsilon $. In
this case, $F(x^{\star })>F(x)$ requires that $\Pi (x^{\star },x)>\Pi (x,x)$%
. We then define an strategy $x^{\star }$ to be evolutionarily stable iff
for all strategies $x\neq x^{\star }$ either

\begin{gather}
\text{1) either }\Pi (x^{\star },x^{\star })-\Pi (x,x^{\star })>0\text{ or if%
}  \notag \\
\text{2) }\Pi (x^{\star },x^{\star })=\Pi (x,x^{\star })\text{ then }\Pi
(x^{\star },x)-\Pi (x,x)>0.  \label{ESS-def}
\end{gather}%
This definition shows that an ESS is a symmetric NE~\cite{Weibull,Samuelson}
satisfying an additional stability property. The stability property ensures
that~\cite{Samuelson} if an ESS establishes itself in a population, it is
able to withstand pressures of mutation and selection. Using a
game-theoretic wording, an ESS is a refinement on the set of symmetric Nash
equilibria and, though being a static solution concept, it describes dynamic
evolutionary situations.

\section{ESS in Prisoner's Dilemma when joint probabilities are factorizable}

We consider game-theoretic solution-concept of an ESS in quantum mechanical
regime by observing that quantum mechanics can make only probabilistic
predictions and any setup for a quantum game must have a probabilistic
description. That is, when a quantum game is constructed using joint
probabilities, even the so-called one-shot game must first be translated
into some appropriate probabilistic version before one considers its quantum
version. This translation permits us, in the following step, to introduce
(quantum mechanical) joint probabilities (that may not be factorizable) and
to find if and how such probabilities can change the outcome of the game.

To achieve this in view of the ESS concept, we consider an EPR-Bohm type
setting~\cite{IqbalCheon} consisting of a bipartite dichotomic physical
system that the two players share to play the game (\ref{matrix}). This
system can be described by the following $16$ joint probabilities $p_{i}$
with $1\leq i\leq 16$:

\begin{equation}
p_{i}=\Pr (\pi _{1},\pi _{2};a,b){\ }\text{with}{\mathrm{\ \ \ }}i=1+\frac{%
(1-\pi _{2})}{2}+2\frac{(1-\pi _{1})}{2}+4(b-1)+8(a-1),  \label{Cerenotation}
\end{equation}%
where $\pi _{1}$ is player $1$'s outcome, that can have a dichotomic value
of $+1$ or $-1$, obtained when s/he plays the strategy $S$ or $S^{\prime }$.
We associate $S\sim 1$ and $S^{\prime }\sim 2$ that then assigns a value for 
$a$. Similarly, $\pi _{2}$ is player $2$'s outcome, that can have a
dichotomic value of $+1$ or $-1$, obtained when s/he plays the strategy $S$
or $S^{\prime }$. The same association $S\sim 1$ and $S^{\prime }\sim 2$
then assigns a value for $b$. For example, the joint probability
corresponding to the situation when player $1$'s outcome $\pi _{1}$ is $+1$
when s/he plays $S^{\prime }$ (i.e. $a=2$), while player $2$'s outcome $\pi
_{2}$ is $-1$ when s/he plays $S$ (i.e. $b=1$), is obtained from (\ref%
{Cerenotation}) as $p_{10}$.

We now define players' payoff relations when they play the game (\ref{matrix}%
) using this (probabilistic) physical system to which the $16$ joint
probabilities (\ref{Cerenotation}) correspond,

\begin{equation}
\Pi _{A,B}(x,y)=\left( 
\begin{array}{c}
x \\ 
1-x%
\end{array}%
\right) ^{T}\left( 
\begin{array}{cc}
\Pi _{A,B}(S,S) & \Pi _{A,B}(S,S^{\prime }) \\ 
\Pi _{A,B}(S^{\prime },S) & \Pi _{A,B}(S^{\prime },S^{\prime })%
\end{array}%
\right) \left( 
\begin{array}{c}
y \\ 
1-y%
\end{array}%
\right) ,  \label{payoffs}
\end{equation}%
where

\begin{eqnarray}
\Pi _{A,B}(S,S) &=&\tsum\nolimits_{i=1}^{4}(a,b)_{i}p_{i},\text{ \ \ \ \ \ }%
\Pi _{A,B}(S,S^{\prime })=\tsum\nolimits_{i=5}^{8}(a,b)_{i-4}p_{i},  \notag
\\
\Pi _{A,B}(S^{\prime },S) &=&\tsum\nolimits_{i=9}^{12}(a,b)_{i-8}p_{i},\text{
\ \ }\Pi _{A,B}(S^{\prime },S^{\prime
})=\tsum\nolimits_{i=13}^{16}(a,b)_{i-12}p_{i}.  \label{payoffs-parts}
\end{eqnarray}%
Here $T$ indicates transpose and $x$ and $y$ are the probabilities,
definable over a large number of runs, with which Alice and Bob choose the
strategies $S$ and $S^{\prime }$, respectively. Joint probabilities are
normalized i.e.

\begin{equation}
\tsum\nolimits_{i=1}^{4}p_{i}=1=\tsum\nolimits_{i=5}^{8}p_{i},\text{ \ \ }%
\tsum\nolimits_{i=9}^{12}p_{i}=1=\tsum\nolimits_{i=13}^{16}p_{i}.
\label{normalization}
\end{equation}

A Nash equilibrium strategy pair $(x^{\star },y^{\star })$ is then obtained
from the inequalities:

\begin{equation}
\Pi _{A}(x^{\star },y^{\star })-\Pi _{A}(x,y^{\star })\geqslant 0,\text{ }%
\Pi _{B}(x^{\star },y^{\star })-\Pi _{B}(x^{\star },y)\geqslant 0,
\label{NE}
\end{equation}%
and a symmetric game, defined by the conditions (\ref{symmetric game}), is
obtained when

\begin{eqnarray}
\Pi _{A}(S,S) &=&\Pi _{B}(S,S),\text{ }\Pi _{A}(S,S^{\prime })=\Pi
_{B}(S^{\prime },S),  \notag \\
\Pi _{A}(S^{\prime },S) &=&\Pi _{B}(S,S^{\prime }),\text{ }\Pi
_{A}(S^{\prime },S^{\prime })=\Pi _{B}(S^{\prime },S^{\prime }).
\label{conds for symmetric game}
\end{eqnarray}%
As it is reported in Ref.~\cite{IqbalCheon}, in case joint probabilities are
factorizable one can find $r,s,r^{\prime },s^{\prime }\in \lbrack 0,1]$ such
that~\cite{IqbalCheon}

\begin{eqnarray}
{\small p}_{1} &{\small =}&{\small rr}^{\prime }{\small ,}\text{ }{\small p}%
_{2}{\small =r(1-r}^{\prime }{\small ),}\text{...}{\small p}_{8}{\small %
=(1-r)(1-s}^{\prime }{\small ),}  \notag \\
{\small p}_{9} &{\small =}&{\small sr}^{\prime }{\small ,}\text{ }{\small p}%
_{10}{\small =s(1-r}^{\prime }{\small ),}\text{...}{\small p}_{16}{\small %
=(1-s)(1-s}^{\prime }{\small ),}  \label{factorizability}
\end{eqnarray}%
and the Nash inequalities (\ref{NE}) are reduced to~\cite{IqbalCheon}

\begin{equation}
(\text{\c{r}}-\text{\c{s}})^{T}\mathcal{A}\left\{ y^{\star }(\text{\c{r}}%
^{\prime }-\text{\c{s}}^{\prime }\mathbf{)+}\text{\c{s}}^{\prime }\right\}
(x^{\star }-x)\geqslant 0,\text{ \ }\left\{ x^{\star }(\text{\c{r}}-\text{%
\c{s}})^{T}+\text{\c{s}}^{T}\right\} \mathcal{B}\mathbf{(}\text{\c{r}}%
^{\prime }\mathbf{-}\text{\c{s}}^{\prime }\mathbf{)}(y^{\star }-y)\geqslant
0,
\end{equation}%
where {\small \c{r}}$=%
\begin{pmatrix}
r \\ 
1-r%
\end{pmatrix}%
,${\small \ \c{s}}$=%
\begin{pmatrix}
s \\ 
1-s%
\end{pmatrix}%
,${\small \ \c{r}}$^{\prime }=%
\begin{pmatrix}
r^{\prime } \\ 
1-r^{\prime }%
\end{pmatrix}%
,${\small \ \c{s}}$^{\prime }=%
\begin{pmatrix}
s^{\prime } \\ 
1-s^{\prime }%
\end{pmatrix}%
$.

When joint probabilities are factorizable, the conditions (\ref{conds for
symmetric game}) to obtain a symmetric game can be shown to reduce to $%
\mathcal{A=B}^{T}$ and the payoff relations (\ref{payoffs}) are then
simplified to

\begin{equation}
\Pi (x,y)=\left( 
\begin{array}{c}
x \\ 
1-x%
\end{array}%
\right) ^{T}\left( 
\begin{array}{cc}
\Pi (S,S) & \Pi (S,S^{\prime }) \\ 
\Pi (S^{\prime },S) & \Pi (S^{\prime },S^{\prime })%
\end{array}%
\right) \left( 
\begin{array}{c}
y \\ 
1-y%
\end{array}%
\right) ,  \label{sym-game-payoffs}
\end{equation}%
where

\begin{eqnarray}
\Pi (S,S) &=&\text{\c{r}}^{T}\mathcal{M}\text{\c{r}}^{\prime },\text{ }\Pi
(S,S^{\prime })=\text{\c{r}}^{T}\mathcal{M}\text{\c{s}}^{\prime },  \notag \\
\Pi (S^{\prime },S) &=&\text{\c{s}}^{T}\mathcal{M}\text{\c{r}}^{\prime }%
\mathbf{,}\text{ }\Pi (S^{\prime },S^{\prime })=\text{\c{s}}^{T}\mathcal{M}%
\text{\c{s}}^{\prime },  \label{sym-game-payoffs-parts}
\end{eqnarray}%
and $\mathcal{M=A=B}^{T}$. The second inequality in (\ref{NE}) is $\Pi
_{B}(x^{\star },y^{\star })-\Pi _{B}(x^{\star },y)\geq 0$ that becomes $\Pi
_{A}(y^{\star },x^{\star })-\Pi _{A}(y,x^{\star })\geq 0$ for a symmetric
game. Comparing it to the first inequality in (\ref{NE}) gives $x^{\star
}=y^{\star }$ and $x=y$ and the definition of a symmetric NE is reduced
simply to $\Pi (x^{\star },x^{\star })-\Pi (x,x^{\star })\geqslant 0$.

Evaluating the two parts of the ESS definition (\ref{ESS-def}) from a
symmetric game payoff relations (\ref{sym-game-payoffs}) we find

\begin{equation}
\Pi (x^{\star },x^{\star })-\Pi (x,x^{\star })=(x^{\star }-x)(x^{\star
}\Delta _{1}+\Delta _{2}),\text{ }\Pi (x^{\star },x)-\Pi (x,x)=(x^{\star
}-x)(x\Delta _{1}+\Delta _{2})  \label{ESS-def-parts}
\end{equation}%
where $\Delta _{1}=(\Pi (S,S)-\Pi (S^{\prime },S)-\Pi (S,S^{\prime })+\Pi
(S^{\prime },S^{\prime })$ and $\Delta _{2}=\Pi (S,S^{\prime })-\Pi
(S^{\prime },S^{\prime })$. Now $\Delta _{1}$ and $\Delta _{2}$ are
evaluated using (\ref{sym-game-payoffs-parts}) as

\begin{equation}
\Delta _{1}=(r-s)(r^{\prime }-s^{\prime })\Omega _{1}\text{ and }\Delta
_{2}=(r-s)(s^{\prime }\Omega _{1}-\Omega _{2}),  \label{sigmas}
\end{equation}%
where $\Omega _{1}=a_{1}-a_{2}-a_{3}+a_{4}$ and $\Omega _{2}=a_{4}-a_{2}$.
Recall that PD is defined by the constraints $a_{3}>a_{1}>a_{4}>a_{2}$ and
we have $\Omega _{2}>0$, which asks for a natural association of the
strategy of defection in PD to the strategy $x^{\star }=0$ played in the
present setting. When both players play this strategy we obtain from the Eq.
(\ref{sym-game-payoffs}) $\Pi (0,0)=\Pi (S^{\prime },S^{\prime })$, which is
the payoff to each player in the classical game when they both defect. With
this association Eqs.~(\ref{ESS-def-parts}) give $\Pi (0,0)-\Pi
(x,0)=-x\Delta _{2}$ and $\Pi (0,x)-\Pi (x,x)=-x(x\Delta _{1}+\Delta _{2})$,
which correspond to the first and second parts of the ESS definition (\ref%
{ESS-def}), respectively. For this strategy if we take

\begin{equation}
s^{\prime }=\Omega _{2}/\Omega _{1},\text{ }r-s=r^{\prime }-s^{\prime }
\label{constraints}
\end{equation}%
then the two parts of the ESS definition are reduced to

\begin{equation}
\Pi (0,0)-\Pi (x,0)=0,\text{ }\Pi (0,x)-\Pi (x,x)=-x^{2}(r-s)^{2}\Omega _{1}.
\label{ESS-factorizable}
\end{equation}

As $\Omega _{1}=\Omega _{2}-\Omega _{3}$, where $\Omega _{3}=(a_{3}-a_{1})$,
for PD both $\Omega _{2},\Omega _{3}>0$. As $\Omega _{2}>0$ and $s^{\prime
}=\Omega _{2}/\Omega _{1}$ is a probability we require the constraint $%
1>\Omega _{2}/\Omega _{1}>0$ so $\Omega _{1}>\Omega _{2}>0$, from which one
obtains $\Omega _{2}>\Omega _{3}$ i.e.

\begin{equation}
a_{4}-a_{2}>a_{3}-a_{1}  \label{extra-req}
\end{equation}%
along with this, of course, we also have $a_{3}>a_{1}>a_{4}>a_{2}$. The
extra requirement (\ref{extra-req}) defines a subset of the games that are
put under the name of a generalized PD. For this game the result (\ref%
{ESS-factorizable}) states that the strategy $x^{\star }=0$ is not an ESS,
though it is a symmetric NE, when joint probabilities are factorizable, in
the sense described by (\ref{factorizability}), and have the constraints (%
\ref{constraints}) imposed on them.

\section{Obtaining the quantum game}

There can be several different possible routes in obtaining a quantum game.
The general idea is to establish correspondence, as a first step, between
classical feature of a physical system and a classical game in the sense
that classical game results because of those features. In the following
step, the classical feature are replaced by quantum feature, while the
obtained correspondence in the first step is retained. One then looks at the
impact which the quantum feature has on the solution/outcome of the game
under consideration. As the mentioned correspondence can be established in
several possible ways, there can be many different routes in obtaining a
quantum game.

To consider ESS in quantum PD we translate playing of this game in terms of
factorizable joint probabilities, which is achieved in the previous Section.
We then find constraints on these probabilities ensuring that the classical
game remains embedded within the quantum game, which is achieved by Eq.~(\ref%
{constraints}). For factorizable joint probabilities the Eqs.~(\ref%
{factorizability}) hold that permit us to translate the constraints (\ref%
{constraints}) in terms of joint probabilities. In the following step, the
joint probabilities are allowed to be non-factorizable, while they continue
to be restricted by the obtained constraints.

Joint probabilities $p_{i}$ become non-factorizable when one cannot find $%
r,s,r^{\prime },s^{\prime }\in \lbrack 0,1]$ such that $p_{i}$ can be
expressed in terms of them i.e.~as given in (\ref{factorizability}). The
same payoff relations (\ref{payoffs}), therefore, correspond to the qunatum
game, whose parts are given by (\ref{payoffs-parts}), and players'
strategies remain exactly the same.

We require that the constraints (\ref{constraints}), when they are
re-expressed using (\ref{factorizability}) in terms of joint probabilities $%
p_{i}$, remain valid while $p_{i}$ are allowed to be non-factorizable. We
notice that Eqs.~(\ref{factorizability}) allow re-expressing the constraints
(\ref{constraints}) in terms of $p_{i}$ as

\begin{equation}
r=p_{1}+p_{2},\text{ }r^{\prime }=p_{1}+p_{3},\text{ }s=p_{9}+p_{10},\text{ }%
s^{\prime }=p_{5}+p_{7},
\end{equation}%
and the constraints (\ref{constraints}) take the form

\begin{equation}
p_{5}+p_{7}=\Omega _{2}/\Omega _{1},\text{ }%
p_{1}+p_{2}-p_{9}-p_{10}=p_{1}+p_{3}-p_{5}-p_{7}.
\label{re-expressed-constraints}
\end{equation}

At this stage we refer to the analysis of joint probabilities in generalized
EPR-Bohm experiments by Cereceda~\cite{Cereceda} reporting that eight out of
sixteen joint probabilities can be eliminated using the normalization
constraints (\ref{normalization}) and the causal communication constraints
given as follows,

\begin{equation}
\begin{array}{cccc}
{\small p}_{1}{\small +p}_{2}{\small =p}_{5}{\small +p}_{6}{\small ,} & 
{\small p}_{1}{\small +p}_{3}{\small =p}_{9}{\small +p}_{11}{\small ,} & 
{\small p}_{9}{\small +p}_{10}{\small =p}_{13}{\small +p}_{14}{\small ,} & 
{\small p}_{5}{\small +p}_{7}{\small =p}_{13}{\small +p}_{15}{\small ,} \\ 
{\small p}_{3}{\small +p}_{4}{\small =p}_{7}{\small +p}_{8}{\small ,} & 
{\small p}_{11}{\small +p}_{12}{\small =p}_{15}{\small +p}_{16}{\small ,} & 
{\small p}_{2}{\small +p}_{4}{\small =p}_{10}{\small +p}_{12}{\small ,} & 
{\small p}_{6}{\small +p}_{8}{\small =p}_{14}{\small +p}_{16}{\small .}%
\end{array}
\label{locality}
\end{equation}

The constraints (\ref{normalization}, \ref{locality}), of course, do hold
for factorizable joint probabilities that are given by Eqs.~(\ref%
{factorizability}). Cereceda expresses probabilities $p_{2},$ $p_{3},$ $%
p_{6},$ $p_{7},$ $p_{10},$ $p_{11},$ $p_{13},$ $p_{16}$ in terms of
probabilities $p_{1},$ $p_{4},$ $p_{5},$ $p_{8},$ $p_{9},$ $p_{12},$ $%
p_{14}, $ $p_{15}$ as

\begin{equation}
\begin{array}{cc}
{\small p}_{2}{\small =(1-p}_{1}{\small -p}_{4}{\small +p}_{5}{\small -p}_{8}%
{\small -p}_{9}{\small +p}_{12}{\small +p}_{14}{\small -p}_{15}{\small )/2,}
& {\small p}_{3}{\small =(1-p}_{1}{\small -p}_{4}{\small -p}_{5}{\small +p}%
_{8}{\small +p}_{9}{\small -p}_{12}{\small -p}_{14}{\small +p}_{15}{\small %
)/2,} \\ 
{\small p}_{6}{\small =(1+p}_{1}{\small -p}_{4}{\small -p}_{5}{\small -p}_{8}%
{\small -p}_{9}{\small +p}_{12}{\small +p}_{14}{\small -p}_{15}{\small )/2,}
& {\small p}_{7}{\small =(1-p}_{1}{\small +p}_{4}{\small -p}_{5}{\small -p}%
_{8}{\small +p}_{9}{\small -p}_{12}{\small -p}_{14}{\small +p}_{15}{\small %
)/2,} \\ 
{\small p}_{10}{\small =(1-p}_{1}{\small +p}_{4}{\small +p}_{5}{\small -p}%
_{8}{\small -p}_{9}{\small -p}_{12}{\small +p}_{14}{\small -p}_{15}{\small %
)/2,} & {\small p}_{11}{\small =(1+p}_{1}{\small -p}_{4}{\small -p}_{5}%
{\small +p}_{8}{\small -p}_{9}{\small -p}_{12}{\small -p}_{14}{\small +p}%
_{15}{\small )/2,} \\ 
{\small p}_{13}{\small =(1-p}_{1}{\small +p}_{4}{\small +p}_{5}{\small -p}%
_{8}{\small +p}_{9}{\small -p}_{12}{\small -p}_{14}{\small -p}_{15}{\small %
)/2,} & {\small p}_{16}{\small =(1+p}_{1}{\small -p}_{4}{\small -p}_{5}%
{\small +p}_{8}{\small -p}_{9}{\small +p}_{12}{\small -p}_{14}{\small -p}%
_{15}{\small )/2,}%
\end{array}
\label{cereceda-analysis}
\end{equation}%
and the payoff relations (\ref{payoffs}) now involve only eight
`independent' probabilities.

\section{ESS in quantum Prisoner's Dilemma}

For the strategy of defection ($x^{\star }=0$) in the quantum game with
payoff relations (\ref{payoffs}), the ESS definition (\ref{ESS-def}) and
Eqs.~(\ref{ESS-def-parts}) give

\begin{equation}
\Pi (0,0)-\Pi (x,0)=x\left\{ \Pi (S^{\prime },S^{\prime })-\Pi (S,S^{\prime
})\right\} ,  \label{QGame-ESS-first}
\end{equation}%
which is equated to zero so that the strategy $x^{\star }=0$ remains a
symmetric NE in the quantum game, as it is the case in the game when joint
probabilities are factorizable, and which is described by Eqs.~(\ref%
{ESS-factorizable}). With setting $\Pi (S^{\prime },S^{\prime })=\Pi
(S,S^{\prime })$ the second part of the ESS definition (\ref{ESS-def}),
which is evaluated in (\ref{ESS-def-parts}), reduces itself to

\begin{equation}
\Pi (0,x)-\Pi (x,x)=x^{2}\left\{ \Pi (S^{\prime },S)-\Pi (S,S)\right\} .
\label{QGame-ESS-second}
\end{equation}

With Cereceda's analysis and using Eqs.~(\ref{cereceda-analysis}), setting $%
\Pi (S,S^{\prime })-\Pi (S^{\prime },S^{\prime })=0$ results in

\begin{equation}
{\small p}_{1}{\small +p}_{5}{\small +p}_{8}{\small +p}_{12}{\small +p}_{14}%
{\small +p}_{15}{\small =1+p}_{4}{\small +p}_{9}{\small ,}\text{ }{\small p}%
_{4}{\small +p}_{5}{\small +p}_{8}{\small +p}_{9}{\small +p}_{14}{\small +p}%
_{15}{\small =1+p}_{1}{\small +p}_{12}{\small ,}  \label{constraints-B}
\end{equation}%
and under the constraints (\ref{constraints-B}) the strategy $x^{\star }=0$
then remains a symmetric NE even for non-factorizable joint probabilities.
Also, using Eqs.~(\ref{cereceda-analysis}) the constraints (\ref%
{re-expressed-constraints}) can be re-expressed in term of `independent
probabilities' as

\begin{equation}
(1-p_{1}+p_{4}+p_{5}-p_{8}+p_{9}-p_{12}-p_{14}+p_{15})=\Omega _{2}/\Omega
_{1},\text{ }p_{5}+p_{12}=p_{8}+p_{9},  \label{constraints-A}
\end{equation}%
which allows us to arbitrarily eliminate probabilities $p_{1}$ and $p_{12}$
from the constraints (\ref{constraints-B}) to re-express them as

\begin{equation}
p_{5}+p_{15}=\Omega _{2}/\Omega _{1},\text{ \ \ }p_{8}+p_{14}=1-\Omega
_{2}/\Omega _{1}.  \label{for p14 & p15}
\end{equation}

Using Eqs.~(\ref{cereceda-analysis}), while considering the strategy $%
x^{\star }=0$ for the second part of the ESS definition the Eq.~(\ref%
{QGame-ESS-second}) becomes

\begin{equation}
\Pi (0,x)-\Pi (x,x)=x^{2}\left\{ \Omega _{3}(p_{1}-p_{9})+\Omega
_{2}(p_{12}-p_{4})\right\} ,
\end{equation}%
which simplifies further when we eliminate $p_{1}$ and $p_{12}$ using (\ref%
{constraints-A}) and afterwards eliminate $p_{14}$ and $p_{15}$ using (\ref%
{constraints-B}) to obtain

\begin{equation}
\Pi (0,x)-\Pi (x,x)=x^{2}(p_{8}+p_{9}-p_{4}-p_{5})\Omega _{1}.
\label{ESS-2nd-f}
\end{equation}%
As $\Omega _{1}>0$, the strategy $x^{\star }=0$ thus becomes an ESS if

\begin{equation}
p_{8}+p_{9}>p_{4}+p_{5},  \label{f-cond}
\end{equation}%
and when joint probabilities $p_{i}$ satisfy constraints (\ref{constraints-A}%
, \ref{constraints-B}), along with the constraints given by normalization
and causal communication.

\section{Discussion}

The game-theoretic solution concept of an ESS is investigated within a
quantization scheme that constructs quantum games from the non-factorizable
property of quantum mechanical joint probabilities. Neither entanglement nor
violation of Bell's inequality ~\cite{Bell,Peres} is used explicitly in this
construction \cite{Ref}.

Eq.~(\ref{ESS-2nd-f}) shows that probabilities $p_{4},$ $p_{5},$ $p_{8},$ $%
p_{9}$ can be taken to be `independent' as, out of the remaining four
probabilities, the probabilities $p_{14}$ and $p_{15}$ are obtained from (%
\ref{for p14 & p15}) and probabilities $p_{1}$ and $p_{12}$ are obtained
from (\ref{constraints-A}). The remaining eight probabilities $p_{2},$ $%
p_{3},$ $p_{6},$ $p_{7},$ $p_{10},$ $p_{11},$ $p_{13},$ $p_{16}$ are then
obtained from (\ref{cereceda-analysis}). The scheme used to obtain a quantum
game assumes that a set of non-factorizable joint probabilities, which
satisfies normalization (\ref{normalization}) and the causal communication
constraint (\ref{locality}) can always be generated by some bipartite
quantum state (pure or mixed) provided that the set does not violate CHSH
form of Bell's inequality beyond Cirel'son's limit~\cite{Cirelson}.

A natural question here is to ask if Bell's inequality is violated by
requiring $p_{8}+p_{9}>p_{4}+p_{5}$, which makes the strategy of defection ($%
x^{\star }=0)$ an ESS. To answer this we consider probabilistic form~\cite%
{Cereceda} of CHSH version of Bell's inequality~\cite{Peres} expressed as $%
-2\leq \Delta \leq 2$ where $\Delta
=2(p_{1}+p_{4}+p_{5}+p_{8}+p_{9}+p_{12}+p_{14}+p_{15}-2)$. We insert values
for $p_{1},$ $p_{12},$ $p_{14},$ $p_{15}$ using (\ref{constraints-A}, \ref%
{for p14 & p15}) to obtain

\begin{equation}
\Delta =2(2p_{4}+p_{9}-1).  \label{CHSH-reduced}
\end{equation}%
Now, comparing (\ref{CHSH-reduced}) to (\ref{f-cond}) shows that the
violation of the CHSH inequality is not essential for the strategy of
defection to be an ESS for a set of non-factorizable probabilities, when for
a factorizable set of probabilities this strategy is non-ESS and a symmetric
NE only.

To summarize, a non-classical solution for an ESS in the quantum PD game has
been shown to emerge due to joint probabilities that are non-factorizable.
An ESS offers a stronger solution concept than a NE and we consider the
situation in which the same NE, consisting of the strategy of defection on
behalf of both players, continues to exist in both the classical and the
quantum versions of the PD game, which correspond to situations of joint
probabilities being factorizable and non-factorizable, respectively. It is
shown that non-factorizable quantum joint probabilities can bring
evolutionary stability to the strategy of defection via the 2nd part of the
ESS definition (\ref{ESS-def}).

\textbf{Acknowledgment:} One of us (AI) is supported at the University of
Adelaide by the Australian Research Council under the Discovery Projects
scheme (Grant No. DP0771453).


\begin{thebibliography}{99}
\bibitem{Ref} As a set of joint probabilities that violates Bell's
inequality must be non-factorizable, a non-classical solution of a game
played in this scheme can emerge even when there is no entanglement and the
quantum state under consideration is separable. This is understandable as a
direct link between violation of Bell's inequality and separability of a
quantum state is established only for pure states via Gisin's theorem~\cite%
{Gisin}. A separable mixed state may still violate a Bell's inequality,
which will correspond to a set of non-factorizable joint probabilities. This
is also consistent with reported results~\cite{CheonTsutsui} showing, for
example, that a quantum game can have a solution in the so-called
\textquotedblleft pseudo-classical domain\textquotedblright , in which
Bell's inequality is not violated. These domains exist between fully
classical and fully quantum domains---where Bell's inequality is violated.
\end{thebibliography}
\end{document}